\documentstyle[a4,11pt,graphicx,epsf,amstex,subfigure]{article}
\newcommand{\be}{\begin{equation}}
\newcommand{\ee}{\end{equation}}
\newcommand{\bd}{\begin{description}}
\newcommand{\ed}{\end{description}}

\newcommand{\Tr}{\mbox{Tr}}
\newcommand{\bea}{\begin{eqnarray}}
\newcommand{\eea}{\end{eqnarray}}

\newcommand{\nn}{\nonumber}

\newcommand{\boldx}{\boldsymbol{x}}
\newcommand{\eqn}[1]{eqn(\ref{#1})}
\newcommand{\Eqn}[1]{Eqn(\ref{#1})}
\newcommand{\eqns}[2]{eqns (\ref{#1}) and (\ref{#2})}
\newcommand{\eqnss}[2]{eqns(\ref{#1}---\ref{#2})}
\newcommand{\fig}[1]{fig~\ref{#1}}
\newcommand{\figs}[2]{figs~\ref{#1} and~\ref{#2}}

\newcommand{\tabss}[2]{tables~\ref{#1}---\ref{#2}}
\makeatletter
\newcommand{\at}{@}
\makeatother
\begin{document}

\title{An Analytic Variational Study of the Mass Spectrum in
    2+1 Dimensional SU(3) Hamiltonian Lattice Gauge Theory}

\author{Jesse Carlsson\thanks{e-mail:{\tt j.carlsson\at
physics.unimelb.edu.au}}, John
A.~L.~McIntosh\thanks{e-mail:{\tt jam\at physics.unimelb.EDU.AU}}, Bruce
H.~J.~McKellar{\thanks{e-mail:{\tt b.mckellar\at physics.unimelb.edu.au}}} \\
and Lloyd C.~L.~Hollenberg\thanks{e-mail:{\tt
L.Hollenberg\at physics.unimelb.edu.au}}\\
\space\\
\em{School of Physics, The University of Melbourne}}

\maketitle
\begin{abstract}
    We calculate the masses of the lowest lying eigenstates of
    improved SU(2) and SU(3) lattice gauge theory in 2+1 dimensions
    using an analytic variational approach.  The ground state is
    approximated by a one plaquette trial state and mass gaps are
    calculated in the symmetric and antisymmetric sectors by
    minimising over a suitable basis of rectangular states.
\end{abstract}

\section{Introduction}

Lattice gauge theory (LGT)~\cite{[1]} is a non-perturbative technique used to regulate 
the ultraviolet divergences of non-abelian
gauge theories. There are two standard formulations of LGT: the
Lagrangian formulation and the Hamiltonian formulation.\\

To date most work on LGT has been carried out in the Lagrangian
formulation for which Monte Carlo techniques are readily applicable. 
This approach involves replacing infinite and continuous space
and time by a finite and discrete lattice of points.  The Hamiltonian
formulation developed by Kogut and Susskind~\cite{[2]} differs by leaving time 
continuous.  In this
framework the problem of gauge theory reduces to a many body problem
and one is interested in calculating ground state and excited state
configurations of the field defined on the lattice. The relation to
continuum physics is realised in the limit in which the lattice
spacing vanishes and the extent of the lattice becomes infinite.\\

Many body techniques, in particular plaquette expansion~\cite{Hollenberg:bp,Hollenberg:pv,McIntosh:jy,Wilson:vf} and the coupled cluster method~\cite{[3],[4],[5],[6]},
have been applied to Hamiltonian LGT for a number of
years.  Not only do such calculations serve as a check on the
more common Monte Carlo techniques of the Lagrangian approach,
they also provide more immediate access to excited states~\cite{[7]}. In
addition, the problem of finite density QCD may be more easily handled
in the Hamiltonian approach~\cite{[8]}.\\

In recent years an improvement program has been undertaken in the
Lagrangian approach. The aim has been to systematically correct the
discretisation errors arising in the lattice Lagrangian and hence
decrease the computational cost of  lattice calculations~\cite{[9]}. More
recently this improvement program has moved to the Hamiltonian
formulation~\cite{[10]}. In a recent paper we demonstrated how improved
Hamiltonians can be constructed~\cite{[11]}. In this paper we explore the use of
improved Hamiltonians in variational calculations of SU(2) and SU(3)
vacuum energy densities, specific heats and exited state energies on a
2+1 dimensional lattice. A key development in this paper is the application of
analytic techniques to SU(3) LGT.\\
   
The outline of this paper is as follows.  In section 2 we introduce a
generating function for SU(3) which allows us perform analytic
calculations of energy densities and specific heats. In section 3 we
calculate the energies of the lowest lying eigenstates of SU(2) and
SU(3) pure gauge theory in 2+1 dimensions and compare the results of
unimproved and improved calculations. Section 4 contains our conclusions.

\section{Vacuum Energy Density and Specific Heat}

\subsection{Introduction}

In this section we calculate vacuum energy densities and specific heats for 
the Kogut Susskind, improved and tadpole improved Hamiltonians. Improved 
Hamiltonians are easily constructed by adding appropriately weighted gauge 
invariant terms to the Kogut Susskind Hamiltonian~\cite{[11]}. For SU($N$) pure 
gauge theory with coupling $g^2$ on a lattice with spacing $a$, we 
define a class of Hamiltonians in terms of the link 
variables $U_{j}(\boldx)$ and electric fields $E_{i}(\boldx)$ (which 
are themselves defined in the appendix), and two parameters $\kappa$ 
and $u_{0}$ which define the particular Hamiltonian in question:
\bea
\tilde{H}(\kappa,u_0) &=& \frac{g^2}{a}\sum_{\boldx,i} \Tr\left[(1-\kappa)E_i(\boldx)^2 + \frac{\kappa}{u_0^2} E_i(\boldx) U_i(\boldx) E_i(\boldx+a \boldsymbol{i}) U^\dagger_i(\boldx)\right]\nn\\
&&  + \frac{2N}{a g^2} \sum_{\boldx, i<j}\left\{(1+4\kappa) P_{ij}(\boldx) -  
\frac{\kappa}{2} \left[R_{ij}(\boldx)+R_{ji}(\boldx)\right]\right\},
\label{ham}
\eea 
where 
\bea
\begin{array}{c}\includegraphics{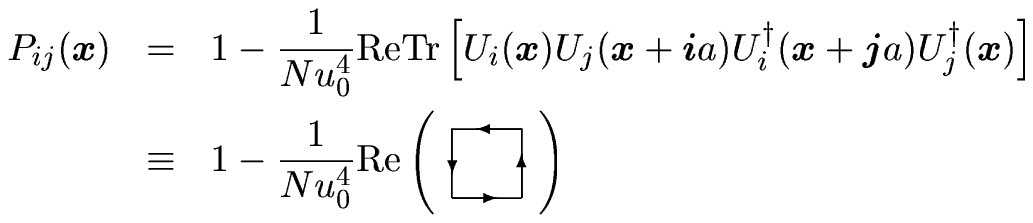}\end{array}
%P_{ij}(\boldx) &=& 1 - \frac{1}{N u_0^4} \Real\Tr\left[U_i(\boldx)U_j(\boldx+\boldsymbol{i}a)
%U^\dagger_i(\boldx+\boldsymbol{j}a) U^\dagger_j(\boldx) \right]\nn\\
%&\equiv & 1 - \frac{1}{N u_0^4} 
%\Real\left(\plaquette\right)
\eea
and
\bea
\begin{array}{c}\includegraphics{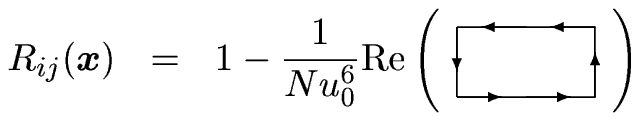}\end{array}
%R_{ij}(\boldx) &=& 1 - \frac{1}{N u_0^6} \Real\left(\rectangleA\right) 
\eea
are the plaquette and rectangle operators respectively.
The Kogut-Susskind and ${\cal O}(a^2)$ classically 
improved Hamiltonians are given by $H_{KS} = \tilde{H}(0,1)$ 
and $H_{I} = \tilde{H}(1/6,1)$ respectively.   
The tadpole improved Hamiltonian is given by $H_{TI} = \tilde{H}(1/6,u_0)$, 
where $u_0$ is now the mean 
link, defined by 
\bea
\begin{array}{c}\includegraphics{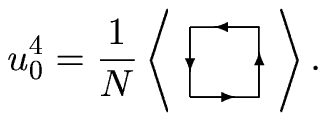}\end{array}
%u_0^4 = \frac{1}{N}\left\langle \plaquette\right\rangle.
\eea  
and evaluated self-consistently, as a function of $g^2$. 
Details of the mean link calculation are given in section~\ref{VEDresults}.\\

We approximate the ground, or (perturbed) vacuum, state, with energy
$E_0$, with
the one plaquette trial state, first suggested by
Greensite~\cite{[12]},
\bea
| \phi_0 \rangle &=& \exp\left[
\frac{c}{2}\sum_p(Z_p+\bar{Z}_p)\right]|0\rangle.
\eea
Here, $|0\rangle $ is the strong coupling vacuum defined by
$E^\alpha_i(\boldx)|0\rangle = 0$ for all $i$, $\boldx$ and $\alpha =
1,2,\ldots,N^2-1$.  The $p$-sum extends over all plaquettes on the
lattice and
\bea
Z_p &=& \Tr\left[U_i(\boldx)U_j(\boldx+\boldsymbol{i}a)
U^\dagger_i(\boldx+\boldsymbol{j}a) U^\dagger_j(\boldx) \right],
\eea
denotes the plaquette labelled by $p$ which joins the lattice sites $\boldx$,
$\boldx+\boldsymbol{i}a$, $\boldx+(\boldsymbol{i}+\boldsymbol{j})a$ and $\boldx+\boldsymbol{j}a$. $\bar{Z}_{p}$ is the 
trace of the link operators in the opposite direction.  For a given
coupling the variational parameter $c$ is fixed by minimising the
vacuum energy density, which, after some algebra making use of
\eqn{simplify} and \eqnss{simplify2}{com4} from the appendix, can be
expressed in terms of the expectation values of plaquettes and
rectangles as follows:

\bea
%\epsilon_0 &=& \frac{a}{N_p} \langle \tilde{H} \rangle \nn\\
%&=& \left[(1-\kappa)\left(\frac{N^2-1}{2\beta}\right) c - \frac{2\beta(1+4\kappa)}
%{Nu_0^4}
%\right]\left\langle \plaquette \right\rangle + \frac{2\kappa \beta}{N u_0^6}
%\left\langle \rectangleA
%\right\rangle.
\begin{array}{c}\includegraphics{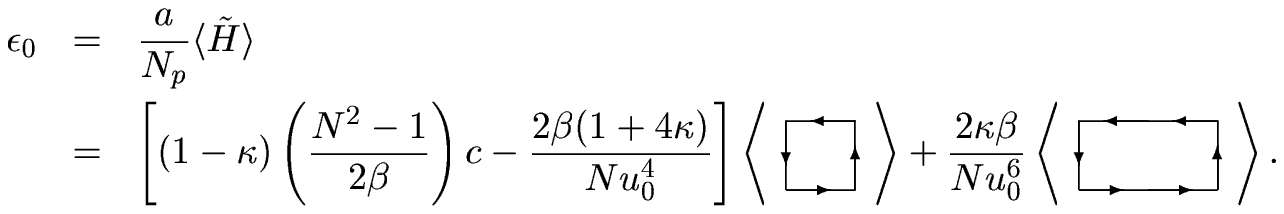}\end{array}
\label{epsilon}
\eea
Here $N_p$ is the number of plaquettes on the lattice and $\beta = N/g^2$ is 
the inverse coupling.

\subsection{Analytic Techniques for SU(2) and SU(3)}

The expression for the SU$(N)$ vaccuum energy density given by
\eqn{epsilon} allows us, in principle, 
to fix the variational parameter $c$ for a
given $\beta$ in any number of dimensions. The difficulty lies in the
calculation of the plaquette and  rectangle expectation values.  Monte
Carlo simulations can be used to calculate these expectation values in
any number of dimensions. However, for the special case of 2+1
dimensions, the  calculation can be carried out analytically. This is
because the  change of variables from links to plaquettes has unit
Jacobian~\cite{[13]}. Consequently, the plaquettes on the two  dimensional
lattice are independent variables, which leads to:
\bea 
\begin{array}{c}\includegraphics{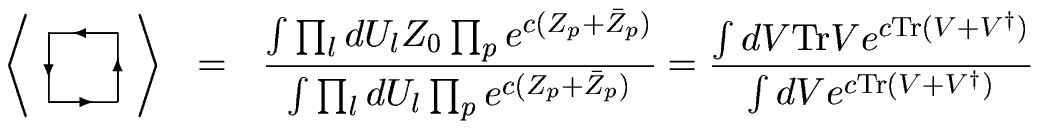}\end{array}
%\left\langle
%\plaquette \right\rangle &=& \frac{ \int\prod_{l}dU_l Z_0 \prod_p
%e^{c(Z_p +\bar{Z}_p)}} {\int\prod_{l}dU_l \prod_p e^{c(Z_p
%+\bar{Z}_p)}}  = \frac{\int dV \Tr V e^{c{\rm Tr}(  V +  V^\dagger)}}
%{\int dV e^{c{\rm Tr}( V +V^\dagger)}} 
\eea 
and 
\bea 
\begin{array}{c}\includegraphics{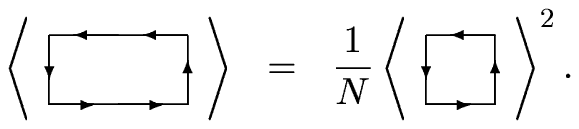}\end{array}
%\left\langle
%\rectangleA \right\rangle &=& \frac{1}{N} \left\langle \plaquette
%\right\rangle^2.
\label{independence}
\eea  
\Eqn{independence} makes use of the independence of plaquettes and \eqn{integrate}. For the case of SU(2), analytic expressions for the 
plaquette expectation value in terms of modified Bessel 
functions ($I_n$) have been used in variational calculations for the last 20 
years:
\bea
\begin{array}{c}\includegraphics{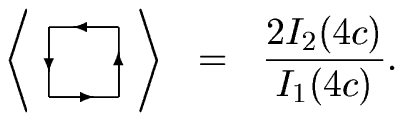}\end{array}
%\left\langle \plaquette \right\rangle &=&  \frac{2 I_2(4c)}{I_1(4c)}.
\eea
We derived the corresponding SU(3) result, which has not been used, to our knowledge, in 
the Hamiltonian approach, and then discovered that it follows simply from a paper of Eriksson et al~\cite{[14]},
in which the SU(3) integral,
\bea
\int dU e^{{\rm Tr}(U M^\dagger +U^\dagger M)},
\eea
is calculated 
for arbitrary $3\times 3$ matrices $M$. Following their analysis, but 
treating $M$ and $M^\dagger$ as independent variables, with $M = c 
1\hspace{-0.8mm}{\rm l}$
and 
$M^\dagger = d 1\hspace{-0.8mm}{\rm l}$, leads to an 
expression for the SU(3) generating function:
\bea
{\cal Y}(c,d) &=&  \int dU e^{{\rm Tr}(cU + dU^\dagger)} \nn\\
&=& \frac{i}{\pi} \oint_\Gamma dz \frac{e^{-(c^3+d^3)/z}}{z (z-cd)^{3/2}} 
J_1\left(\frac{2}{z}(z-cd)^{3/2}\right). 
\eea 
Here $J_1$ is the first Bessel function and $\Gamma$ is a closed contour in the complex plane including the pole at $z=0$ but excluding the pole at $z=cd$. To evaluate this contour integral we expand the integrand in power series about the pole at $z=0$ and use Cauchy's integral theorem to eventually obtain,
\bea
{\cal Y}(c,d) &=& 2 \sum_{k=0}^\infty \frac{1}{(k+1)!(k+2)!}
\sum_{l=0}^k\left(\!\begin{array}{c} 3k+3 \\ k-l \end{array}\!\right)
\frac{1}{l!}(cd)^{k-l}(c^3+d^3)^l. \label{Y}
\eea
This generating functional is extremely useful for Hamiltonian lattice 
calculations\footnote{The SU($N$) generalisation of this result, which
we have recently obtained, will be reported in a
future publication.}. In principle, it allows us to investigate 2+1 dimensional pure SU(3) gauge theory analytically. The calculation of various matrix elements 
for all couplings, reduces to a exercise in differentiation. For example: 
\bea
\begin{array}{c}
\includegraphics{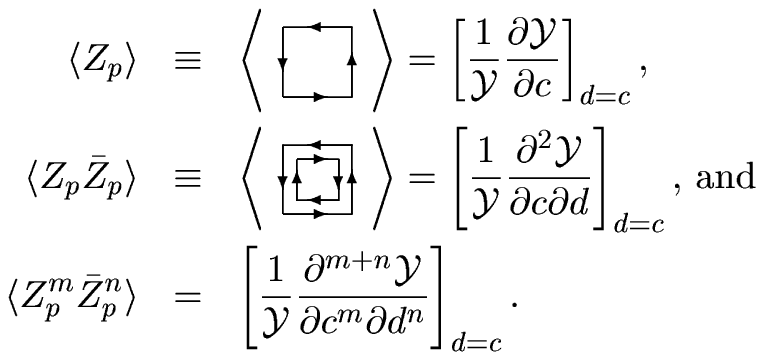}
\end{array}
%\langle Z_p \rangle &\equiv& \left\langle \plaquette \right\rangle 
%= \left[\frac{1}{{\cal Y}}\frac{\partial{\cal Y} }{\partial c}\right]_{d=c}, \nn\\
%\langle Z_p \bar{Z}_p \rangle &\equiv& \left\langle \graphdd \right\rangle 
%= \left[\frac{1}{{\cal Y}}\frac{\partial^2{\cal Y}}
%{\partial c \partial d}\right]_{d=c} \mbox{, and}
%\nn\\
%\langle Z_p^m \bar{Z}_p^n \rangle 
%&=& \left[\frac{1}{{\cal Y}}\frac{\partial^{m+n}{\cal Y} }
%{\partial c^m \partial 
%d^n}\right]_{d=c}.
\eea
As a check on this series 
we calculate the strong coupling limit ($c=0$) of $\langle Z_p^m \bar{Z}_p^n\rangle$. By differentiating \eqn{Y}, we observe that the only non-zero strong coupling matrix elements occur when $n+2m \equiv 0 \,{\rm mod}\, 3$. For this case we have,
\bea
\langle Z_p^m \bar{Z}_p^n\rangle_{c=0} 
&\!\!\!=\!\!\!&
\sum_{k} 
\left(\!\! 
\begin{array}{c}3k\!+\!3 \\ n\!+\!m\!+\!3 \end{array}
\!\!\right)
\frac{2 n! m! }{(k+1)!(k+2)!\left(\frac{n+2m}{3}\!-\!k\right)!
\left(\frac{m+2n}{3}\!-\!k\right)!}.
\label{sc}
\eea
Here the sum runs over all integers $\frac{n+m}{3}\le k \le {\rm min}\left(\frac{n+2m}{3},\frac{m+2n}{3}\right)$.
This strong coupling result has an equivalent combinatoric interpretation as 
the number of times the singlet representation appears in the direct product of $m$ 3 and $n$ $\bar{3}$ representations~\cite{[15]}. A simple Mathematica 
code verifies the agreement of the two approaches.\\
 
\begin{figure}
\centering
\includegraphics[width=10cm]{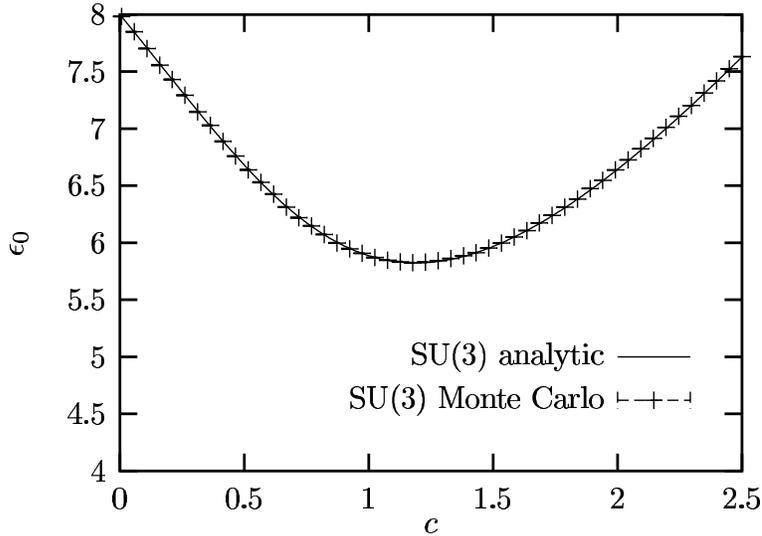}
\caption{Analytic and Monte Carlo calculations of the SU(3) vacuum
energy density in 2+1 dimensions for $\beta = 4.0$.  It should be noted that the
errors in the Monte Carlo calculation are smaller than the plot
points.}
\label{cmontecarlo}   
\end{figure}

A demonstration of the accuracy of the series away from the strong
coupling limit is provided by a Monte Carlo calculation of the vacuum
energy density at various couplings.  The example of $\beta = 4.0$ is
shown in \fig{cmontecarlo} and indicates close agreement between the
Monte Carlo and analytic calculations.  

\subsection{Results}
\label{VEDresults}

To calculate the energy density and variational parameter as functions
of $\beta$ we proceed as follows. For the Kogut-Susskind and
classically improved cases, we simply minimise $\epsilon_0$ for a given
value of $\beta$. The value of $c$ at which $\epsilon_0$ takes its minimum
defines $c$ as a function of $\beta$.  \\

The tadpole improved calculation is more complicated.
This is because the mean plaquette, from which
$u_0$ is calculated, depends on the variational state $|\phi_0\rangle$. 
To complicate
matters, the variational state in turn depends on the energy density
and hence $u_0$. Such interdependence suggests the use of the
following iterative approach. For a given $\beta$ and initial
approximation of $u_0$ we minimise the energy density to fix the
variational state $|\phi_0\rangle $. Using this trial state, we update
$u_0$ for use in the next iteration. This process is iterated until
convergence is achieved, typically after only a few iterations.  \\

\begin{figure}
\centering
\subfigure[SU(2)] % caption for subfigure a
                     {
                         \label{esu2}
                         \includegraphics[width=7cm]{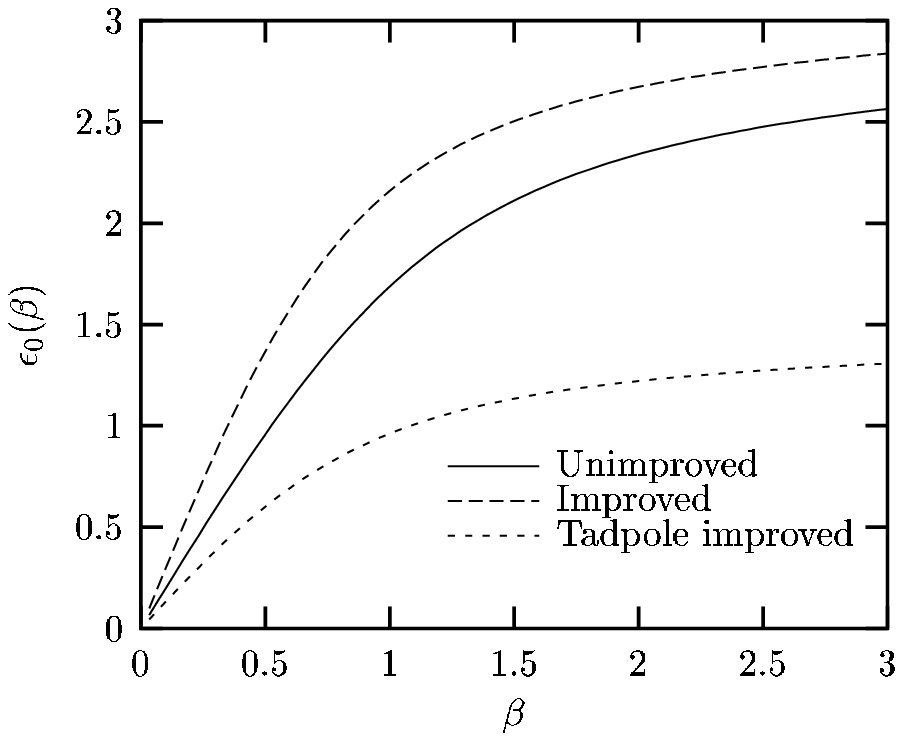}
                     }                   
 \subfigure[SU(3)] % caption for subfigure a
                     {
                         \label{esu3}
                         \includegraphics[width=7cm]{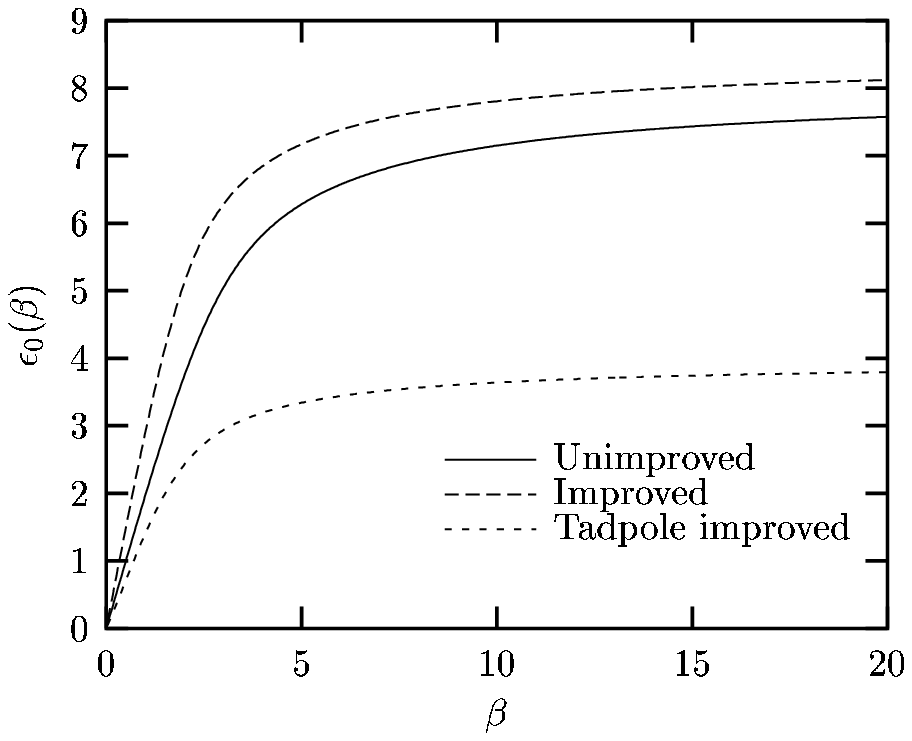}
                     }
\caption{Analytic calculation of the unimproved, improved and tadpole improved vacuum energy density in 2+1 dimensions for SU(2) and SU(3).}
\label{edens}
\end{figure}

\begin{figure}
\centering
\subfigure[SU(2)] % caption for subfigure a
                     {
                         \label{csu2}
                       \includegraphics[width=7cm]{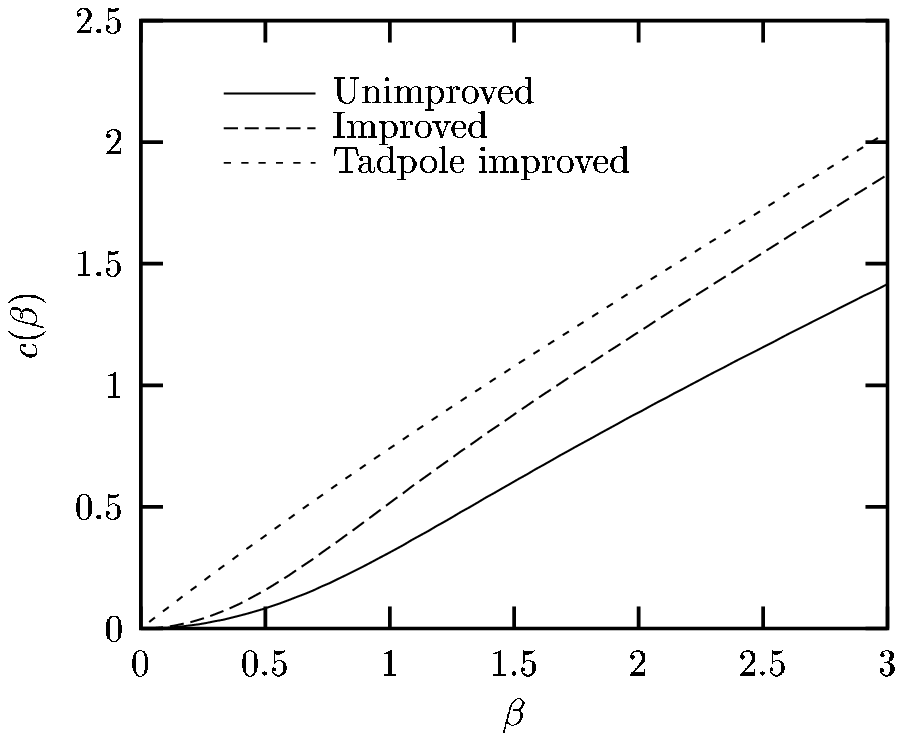}
                     } \hspace{0.25cm}                   
 \subfigure[SU(3)] % caption for subfigure a
                     {
                         \label{csu3}
                         \includegraphics[width=7cm]{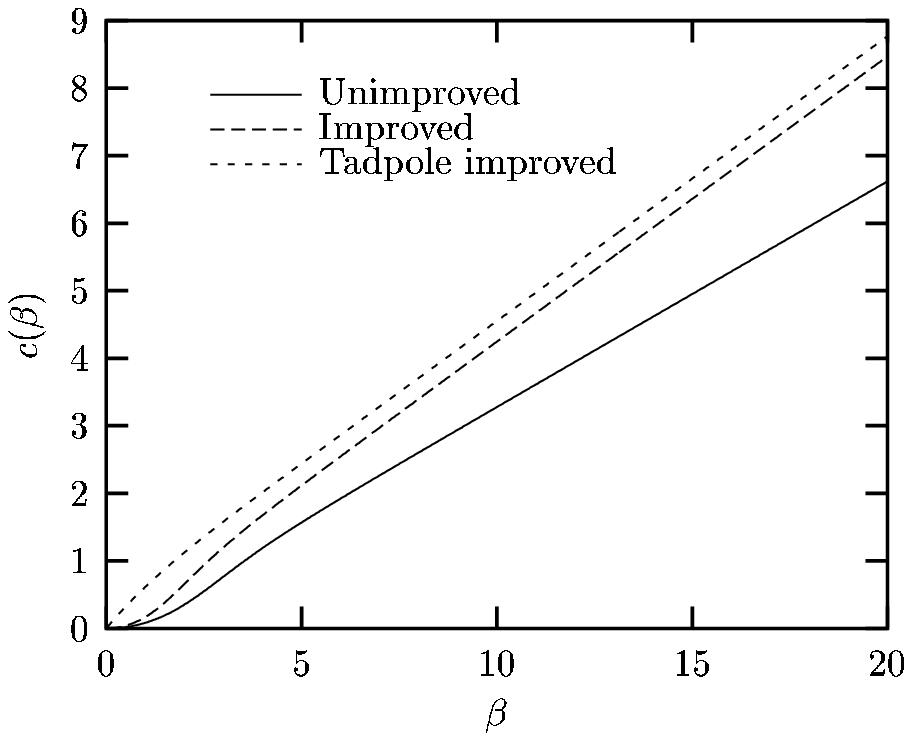}
                     }
\caption{Analytic calculation of the unimproved, improved and tadpole improved variational parameter in 2+1 dimensions for SU(2) and SU(3).}
\label{cofbeta}
\end{figure}

The
results of the Kogut-Susskind, classically improved and tadpole
improved  SU(2) and SU(3) vacuum energy density calculations are shown
in \fig{edens}. The corresponding variational parameters $c(\beta)$
 are shown in \fig{cofbeta}. 
The correct strong and weak coupling behavior is observed in each case.
The differing gradients in the weak coupling limit highlight the fact that when using an improved Hamiltonian one is using a different renormalisation scheme to the unimproved case.\\


\begin{figure}
\centering

\includegraphics[width=10cm]{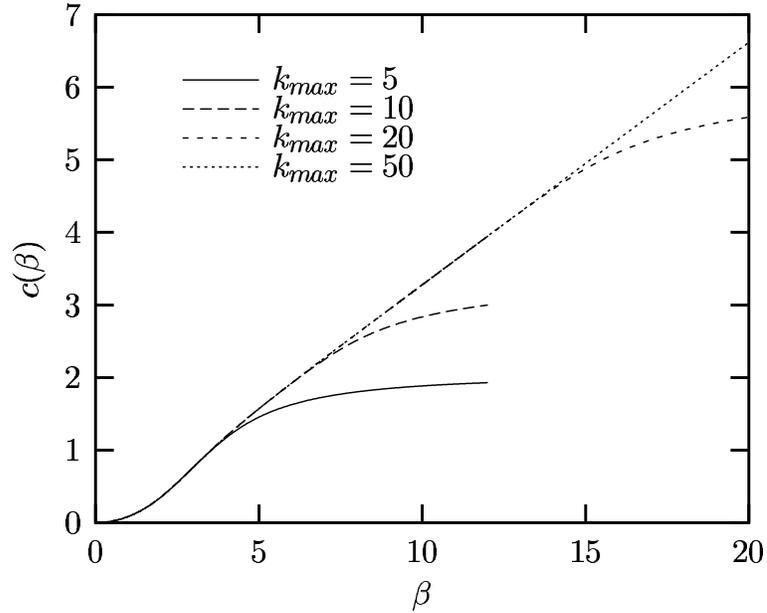}

\caption{Analytic calculation of the unimproved SU(3) variational parameter in 2+1 dimensions, truncating the $k$-sum of ${\cal Y}(c,d)$ at $k=k_{max}$.} 
\label{cconvergence}  
\end{figure}

In practice the SU(3) generating function is truncated. 
The dependence of the variational parameter on various truncations of the $k$-sum in \eqn{Y} 
is shown in \fig{cconvergence}. We see that convergence is achieved up to $\beta \approx 13$ when keeping 20 terms. Further calculations show that when keeping 50 terms convergence up to $\beta \approx 30$ is achieved.\\

\begin{figure}
\centering
\subfigure[SU(2)] % caption for subfigure a
                     {
                         \label{cvsu2}
                         \includegraphics[width=7cm]{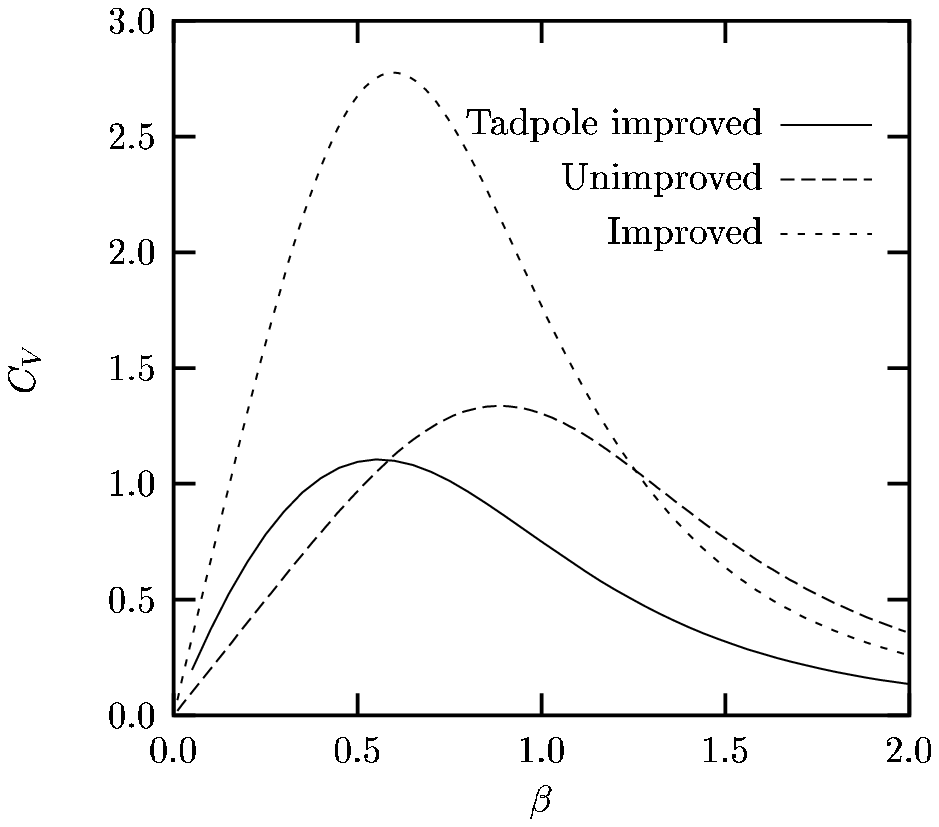}
                     }                   
 \subfigure[SU(3)] % caption for subfigure a
                     {
                         \label{cvsu3}
                         \includegraphics[width=7cm]{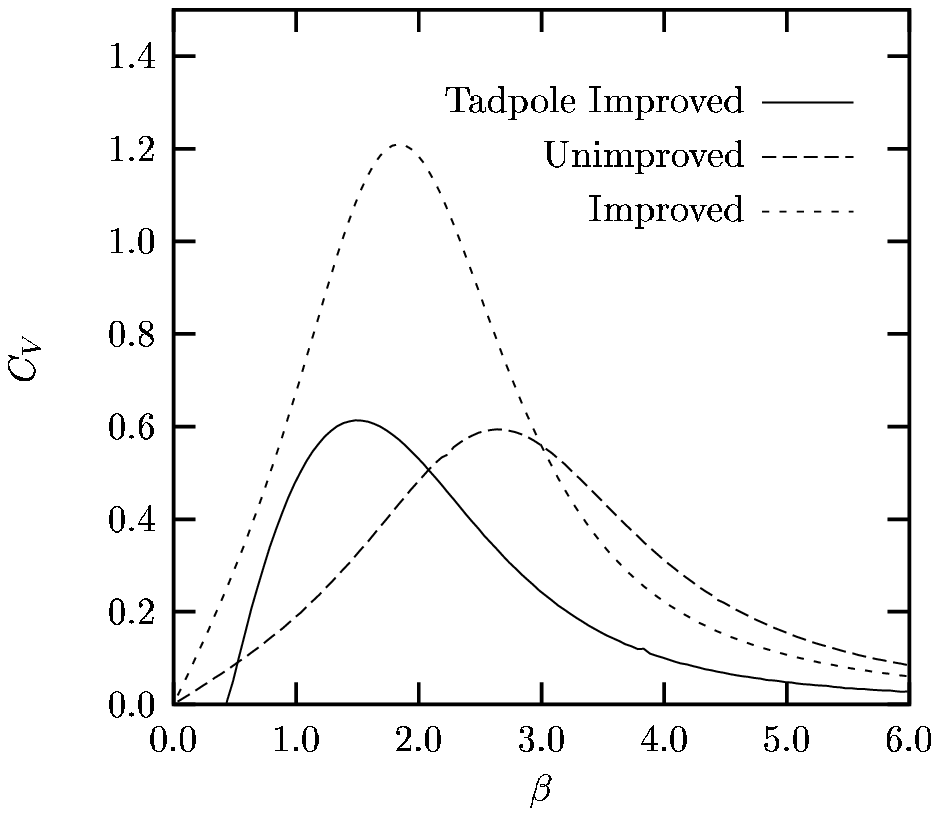}
                     }
\caption{The unimproved, improved and tadpole improved specific heat in 2+1 dimensions for SU(2) and SU(3).}
\label{cv}
\end{figure}

In addition to the vacuum energy density we can also calculate the 
specific heat:
\bea
C_V= -\frac{\partial^2 \epsilon_0}{\partial \beta^2}.
\eea
The results are shown in \fig{cv}. We note that the location of the peak indicates the region of transition from strong to weak coupling. For an improved calculation one would expect the peak to be located at a smaller $\beta$ (corresponding to a larger coupling) than for the equivalent unimproved calculation. We see that this is indeed the case for both SU(2) and SU(3), with the tadpole improved Hamiltonian demonstrating the largest degree of improvement.    

\section{Mass Gaps}

\subsection{Introduction}

Having fixed the chosen trial vacuum $|\phi_0\rangle$, in this section we
 turn to investigating excited states. 
Our aim is to calculate the lowest lying energy eigenstates
of the Hamiltonians described by \eqn{ham} for both SU(2) and SU(3). \\

We follow Arisue~\cite{[16]} and expand the excited state $|\phi_1\rangle$ in the basis consisting of all rectangular Wilson loops $\{|l\rangle\}_{l=1}^{L_{max}}$ that fit in a given square whose side length $L_{max}$
defines the order of the calculation. In order to ensure the orthogonality of $|\phi_0\rangle$ and $|\phi_1\rangle$ we parameterise the excited state as follows
\bea
|\phi_1\rangle &=& \sum_{n,m=1}^{L_{max}} s_{l}|n,m\rangle= \sum_{l=1}^{L_{max}^2} s_{l}|l\rangle, 
\eea
with,
\bea
 |l\rangle &=&  \sum_{\boldx} \left[W_l(\boldx) - \langle W_l(\boldx) 
\rangle\right]|\phi_0\rangle.
\eea
Here $\langle W_l(\boldx) \rangle$ is the expectation value of
$W_l(\boldx)$ with respect to the ground state $|\phi_0\rangle$ and
the convenient label $l=(n-1)L_{max}+m$ has been defined to label the
$n\times m$ rectangular state.  We define the particular form of
$W_l(\boldx)$ to reflect the symmetry of the sector we wish to
consider.  For SU(3) we take $W_l(\boldx)= \Tr[w_l(\boldx)\pm
w^\dagger_l(\boldx)]$ for the symmetric sector and antisymmetric
sector.  To avoid over-decorated equations, the particular 
$W_{l}(\boldx)$ in use is to be deduced from the context.
Here $w_{l}(\boldx)$ is the rectangular Wilson loop joining
the lattice sites $\boldx$, $\boldx+na\boldsymbol{i}$,
$\boldx+na\boldsymbol{i}+ma\boldsymbol{j}$ and
$\boldx+na\boldsymbol{j}$, with
\bea
n = \left[ \frac{l-1}{L_{max}}\right]+1 \quad{\rm and}\quad m = l- L_{max}\left[ \frac{l-1}{L_{max}}\right] .
\eea
Here $[k]$ denotes the integer part of $k$.
In order to calculate excited state energies we minimise the mass gap (the difference between the excited state and ground state energy) over the basis defined by a particular order $L_{max}$. To do this we again follow Arisue~\cite{[16]} and define the matrices
\bea
N_{l l'} &=& \frac{1}{N_p}\langle l|\tilde{H} - E_0|l'\rangle
\eea
and
\bea
D_{l l'} &=& \frac{1}{N_p}\langle l|l'\rangle 
= \sum_{\boldx}\left[\langle W^\dagger_l(\boldx) W_{l'}(\boldsymbol{0})
\rangle  - \langle W_l(\boldx) \rangle^\ast 
\langle W_{l'}(\boldsymbol{0}) \rangle\right].     
\label{d}
\eea 
Extending the calculation to the general improved Hamiltonian $\tilde{H}(\kappa,u_0)$ and making use of \eqnss{com1}{com4} from the appendix gives
\bea
N_{l l'} &=& -\frac{g^2}{2a}\sum_{i,\boldx}\sum_{\boldx'}\Bigg\{
(1-\kappa) \left\langle \left[E^\alpha_i(\boldx),W^\dagger_{l}(\boldx') \right]\left[E^\alpha_i(\boldx),W_{l'}(\boldsymbol{0})\right]\right\rangle \nn\\
&& \hspace{2cm}+ \frac{\kappa}{u_0^2}\left\langle \left[E^\alpha_i(\boldx),W^\dagger_{l}(\boldx')\right]\left[\tilde{E}^\alpha_i(\boldx+a\boldsymbol{i}),W_{l'}(\boldsymbol{0})\right]\right\rangle\Bigg\}.
\label{N}
\eea 
To minimise the mass gap over a basis of a given order we make use of following
 diagonalisation technique~\cite{[17]}. We first diagonalise $D$, with 
\bea
S^\dagger D S = V^{2}, \label{diag}
\eea
where $V$ is diagonal.
The $n$-th lowest eigenvalue of the modified Hamiltonian
\bea
H' = V S^\dagger N S V , 
\eea  
then gives an upper bound for the mass gap corresponding to the 
$n$-th lowest eigenvalue of the Hamiltonian, $\Delta m_n$.\\

\subsection{Calculating Matrix Elements}

Having described the minimisation process
we now detail the calculation of
 the matrix elements $N_{l l'}$ and $D_{l l'}$. Our aim is to reduce $N_{ll'}$ and $D_{ll'}$ to polynomials of one plaquette matrix elements.
This, again has been done for the case of SU(2) by Arisue~\cite{[16]}. Here we retrace his calculations for the case of SU(3) and extend them to incorporate improved Hamiltonians. We start with $D_{l l'}$.\\

Taking elementary plaquettes as our independent variables, it is easy to show that the only non-zero contributions to $D_{l l'}$ occur when the rectangles $l$ and $l'$ overlap. As an example of a contribution to $D_{l l'}$, consider $\Delta D_{l l'}$; the case where $N_{l\cap l'}$ plaquettes are contained by both rectangles (these are the overlap plaquettes) and $N_{l}$ plaquettes are contained by the rectangle $l$.
In order to calculate such matrix elements we rely heavily on the orthogonality properties of SU$(N)$ characters. We are interested in calculating SU$(N)$ 
integrals of the form
\bea
\int dU_p e^{S(U_p)}\chi_r(U_p V), 
\label{interest}
\eea 
where $U_p$ is a SU$(N)$ plaquette variable and $V$ is a product of any number of plaquettes not including $U_p$. Here $\chi_r(U)$ denotes the character corresponding to the representation $r$. For SU(2), $r = 0,1/2,3/2,\ldots$ and for SU(3), $r= (\lambda,\mu)$ where $\lambda$ denotes the number of boxes in the first row of the Young's Tableaux describing the representation and $\mu$ is the number of boxes in the second row. Similarly, for SU$(N)$, $r=(r_1,r_2,\ldots,r_{N-1})$.\\

Performing a character expansion of the exponent in eq.~\ref{interest} gives:
\bea
\int dU_p e^{S(U_p)} \chi_{r}(U_p V) &=& \sum_{r'} \int dU_p
c_{r'} \chi_{r'}(U_p)\chi_{r}(U_p V).
\eea
This is simply a generalisation of a Fourier expansion.
Here, the coefficient $c_{r'}$ is given by:
\bea
c_{r'} = \int dU_p \chi_{r'}(U_p) e^{S(U_p)}.
\eea
Now using the orthogonality relation,
\bea
\int dU_p
\chi_{r'}(U_p V) \chi_{r}(U_p) = \frac{1}{d_r} \delta_{r'r} \chi_r(V),
\eea
where $d_r$ is the dimension of the representation $r$, we obtain:
\bea
\int dU_p e^{S(U_p)} \chi_{r}(U_p V) &=& 
\frac{1}{d_r} \chi_r(V)\int dU_p \chi_{r}(U_p) e^{S(U_p)}.
\label{integrate}
\eea
This result allows us to integrate out a single plaquette from an extended Wilson loop in a given representation $r$. To complete the calculation we need to relate characters to traces of group elements. This can be done using Weyl's character formula~\cite{[18]}. For SU(3) the dimensions and characters corresponding to the first few representations are given by:
\bea
\begin{array}{rclcrcl}
 \chi_{10}(U) &=& \Tr U &\qquad& d_{10} &=& 3\\ 
\vspace{0.1cm}
\chi_{11}(U) &=& \displaystyle
\frac{1}{2}\left[(\Tr U)^2-\Tr(U^2)\right]= \Tr U^\dagger
&\qquad & d_{11} &=& 3\\ 
\vspace{0.1cm}
\chi_{20}(U) &=& \displaystyle
\frac{1}{2}\left[(\Tr U)^2+\Tr(U^2)\right]= (\Tr U)^2-
\Tr U^\dagger  &\qquad & d_{20} &=& 6\\
 \chi_{21}(U) &=& \displaystyle
\frac{1}{3}\left[(\Tr U)^3-\Tr(U^3)\right]= \Tr U \Tr U^\dagger -1 & \qquad & d_{21} &=& 8
\end{array}
\eea     
Here we have made use of the SU(3) identity
\bea
\Tr (U^2) = (\Tr U)^2 - 2\Tr U^\dagger. 
\eea
Applying \eqnss{integrate}{d} allows us to obtain analytic expressions for each contribution to $D_{ll'}$. For the case of $\Delta D_{l l'}$ described earlier, we have 
\bea
\Delta D_{l l'} &=& \frac{2}{3} F_{Z}(N_l+ N_{l'} - 2 N_{l\cap l'})
\left[F_{Z^2}(N_{l\cap l'}) +F_{Z\!\bar{Z}}(N_{l\cap l'}) \right] 
-4 F_{Z}(N_l)F_{Z}(N_{l'}),
\eea
where the character integrals are given by:
\bea
F_{Z}(n) &=& 
% \left\langle
% \plaquette \right\rangle \hspace{-0.98cm}n\hspace{0.7cm} = 
\left(\frac{1}{3}\right)^{n-1} 
\langle Z_p \rangle^n , \\
F_{Z^2}(n) &=& 
%\left\langle\graphd\right\rangle\hspace{-0.98cm}n \hspace{0.7cm}
%=
 \left(\frac{1}{6}\right)^{n-1}\langle Z_p^2
-\bar{Z}_p \rangle^n  +
 \left(\frac{1}{3}\right)^{n-1}\langle \bar{Z}_p \rangle^n \hspace{0.1cm}\mbox{, and}\\
F_{Z\!\bar{Z}}(n) &=& 
%\left\langle\graphdd\right\rangle\hspace{-0.98cm}n \hspace{0.7cm} =
1 +
\left(\frac{1}{8}\right)^{n-1} \left(\langle Z_p \bar{Z}_p
\rangle - 1 \right)^n.
\eea

We now move on to the calculation of $N_{ll'}$. It is easy to show that the only non-zero contributions occur when there is at least one common link and an overlap between the rectangles. The improvement term (the second term in \eqn{N}) only contributes when the two rectangles share at least two links in a given direction. Consider the contribution $\Delta N_{l l'}$ to $N_{ll'}$ in which there are $L_1$ common links and $L_2$ common strings of two links in a given direction. Again we suppose $N_l$ plaquettes are enclosed by rectangle $l$ and that there are $N_{l\cap\l'}$ common plaquettes. 
Making use of \eqn{integrate} and \eqnss{com1}{com4} from the appendix 
we obtain
\bea
\Delta N_{ll'} = \frac{L}{3} F_Z(N_l\!+\!N_{l'}\!-\!2N_{l\cap l'})\left[ 
\frac{2}{3}F_{Z^2}(N_{l\cap l'})
\!-\!2 F_{Z}(N_{l\cap l'})
\!-\! 3 \!+\! \frac{1}{3} F_{Z\!\bar{Z}}(N_{l\cap l'})\right],
\eea
with
\bea
L = (1-\kappa)L_1 + \frac{\kappa}{u_0^2}L_2.
\eea
Having calculated individual contributions to $D_{l l'}$ and $N_{l l'}$, the completion of the calculation requires counting the possible overlaps of a given type. The approach described in this section is easily extended to SU($N$). However, analytic expressions for the character integrals are not available for this case. 

\subsection{Results}

For each SU(3) calculation we have kept 80 terms in the $k$-sum of \eqn{Y} giving convergence up to $\beta = 50$. The generation of $N_{ll'}$ and $D_{ll'}$  and implementation of the minimisation process was accomplished with a Mathematica code. \\

\begin{figure}
\centering
\subfigure[SU(2)] % caption for subfigure a
                     {
                         \label{su2conv}
                         \includegraphics[width=7cm]{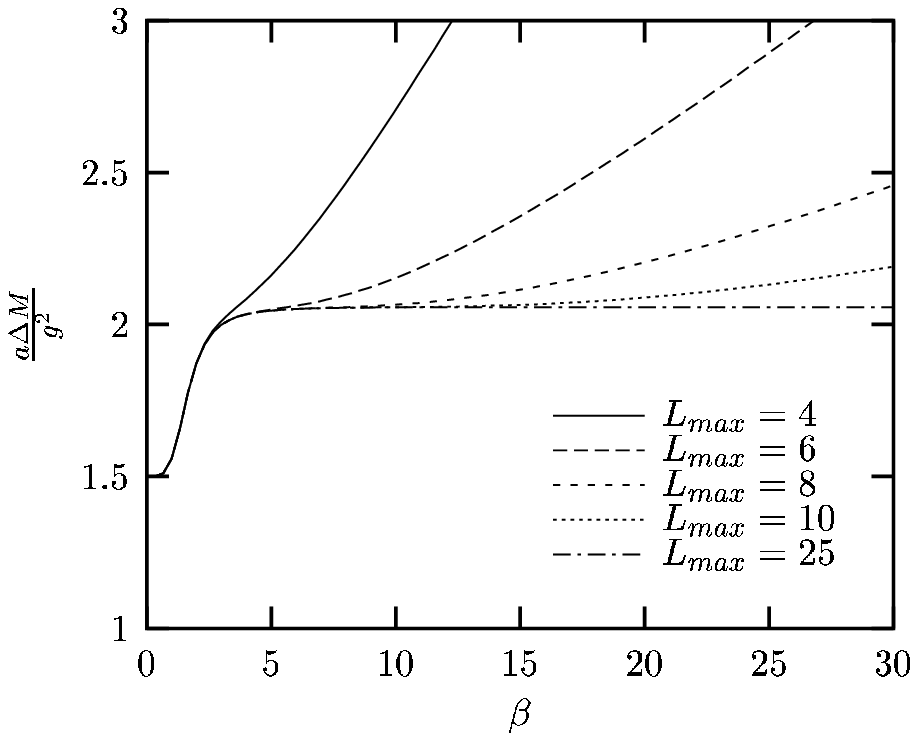}
                     }                   
 \subfigure[SU(3)] % caption for subfigure a
                     {
                         \label{su3conv}
                         \includegraphics[width=7cm]{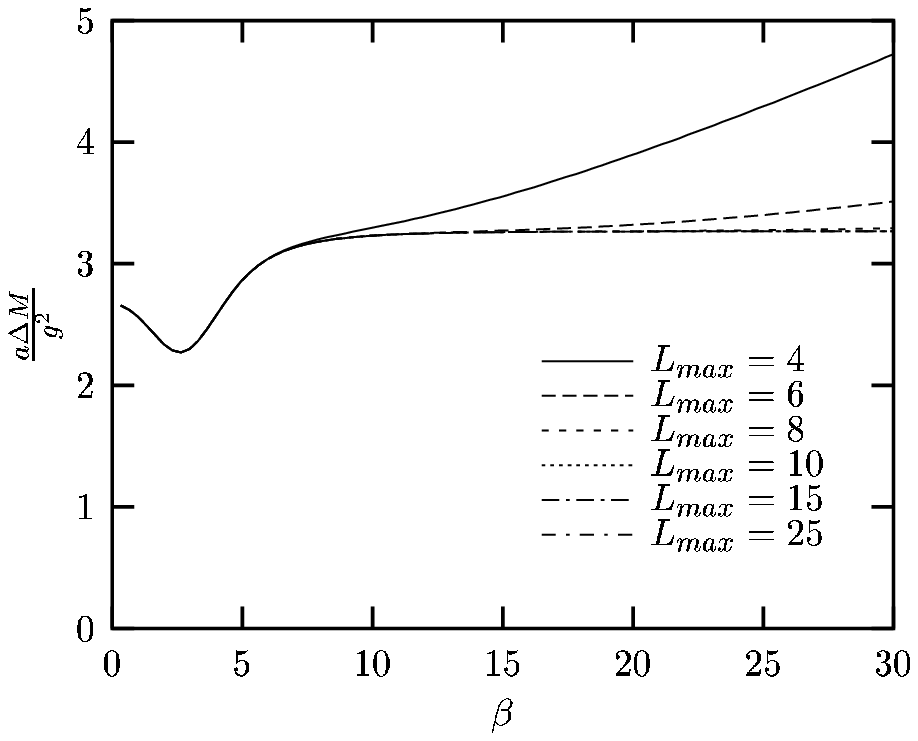}
                     }\caption{The unimproved 2+1 dimensional 
                     symmetric massgaps for SU(2) and SU(3).}
\label{mgconv}
\end{figure}

For the case of 2+1 dimensions we expect $\Delta m/g^2$ to become constant in the scaling region. 
The convergence of the massgaps with $L_{max}$ for the cases of SU(2) and SU(3) is illustrated in \figs{su2conv}{su3conv}. We notice that for the case of SU(3) only small improvements to scaling are gained by extending the calculation beyond order 8. This suggests that a more complicated basis (including, for example, loops covered more than once) is required to simulate SU(3) excited states than for the case of SU(2).\\

\begin{figure}
\centering
\subfigure[Symmetric SU(2) massgap] % caption for subfigure a
                     {
                         \label{su2sym}
			\includegraphics[width=7cm]{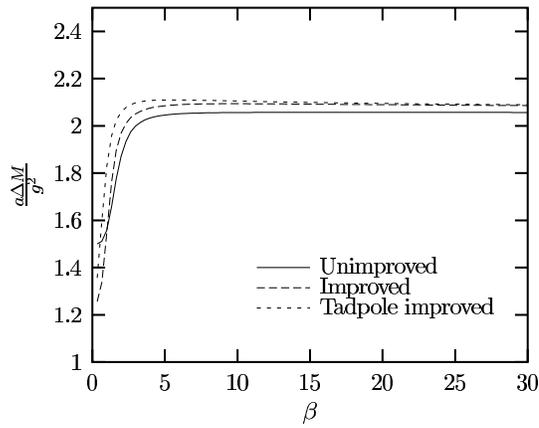}
                     }                   
 \subfigure[Symmetric SU(3) massgap] % caption for subfigure a
                     {
                         \label{su3sym}
	\includegraphics[width=7cm]{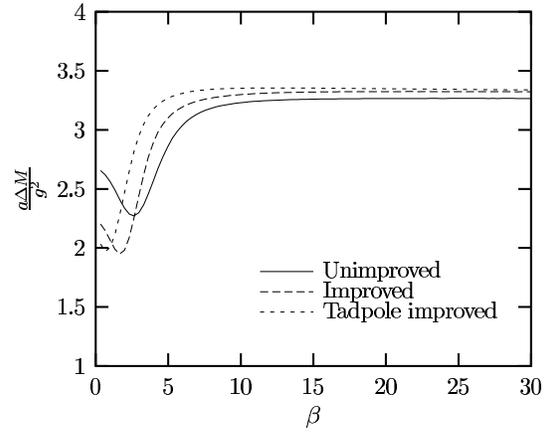}
                     }\\
\subfigure[Antisymmetric SU(3) massgap] % caption for subfigure a
                     {
                         \label{su3asym}
			\includegraphics[width=10cm]{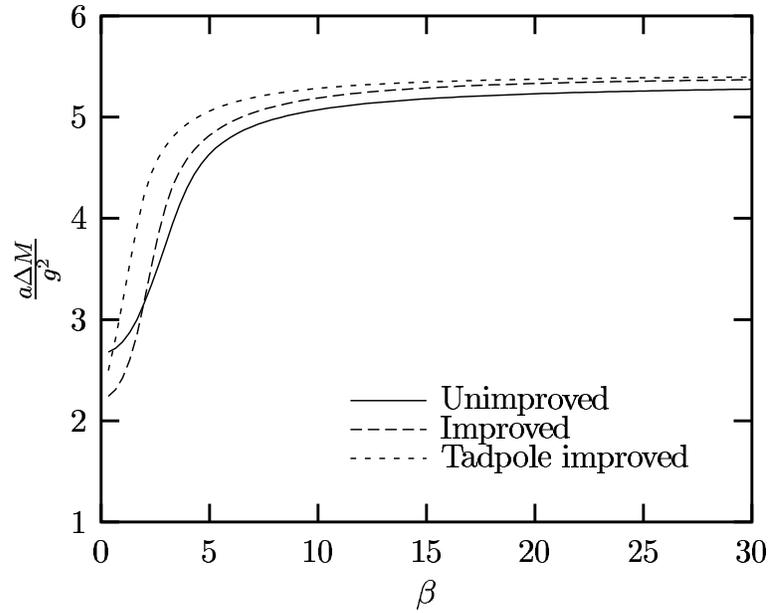}
                     }
\caption{2+1 dimensional  massgaps for SU(2) and SU(3) with $L_{max}=25$.}
\label{mgcompare}
\end{figure}

From \fig{mgcompare}, we see that $\Delta m^S_1/g^2$ is approximated well by a constant in very large scaling regions for the lowest lying eigenstates for both SU(2) and SU(3). The scaling behavior becomes significantly worse for the antisymmetric sector and for higher energy eigenvalues. This is because the simplistic form of our excited state wave function is not sufficient to reproduce the plaquette correlations required to simulate these higher order states. One would expect the simulation of higher order eigenstates to improve by including more complicated loops in our expansion basis or by using a more complicated ground state. The continuum limit results of excited states for SU(2) and SU(3) results are given in \tabss{mgg}{mgmgmg}.\\

For the unimproved SU(2) case, the lowest two eigenstates agree precisely with 
the calculation of Arisue~\cite{[16]}. This serves as a check on our counting in calculating the possible overlaps of excited states. Our calculation is in disagreement with that of Arisue at the 3rd eigenstate. It appears that our 4th eigenstate corresponds to Arisue's 3rd and that our third eigenstate does not appear in his results. The reasons for this are not clear. \\

The results for the SU(3) symmetric massgap (in units of $g^2$)
are to be compared to calculations by Luo and Chen $2.15 \pm 0.06$~\cite{[19]}, Samuel
$1.84 \pm 0.46$~\cite{[20]} and Teper $2.40 \pm 0.02$~\cite{[21]}. It should be stressed that although our result of $3.26520 \pm 0.00009$ is considerably higher, it provides a precise upper upper-bound on the exact result. By including more complicated loops in the expansion basis one would expect to reduce this upper bound. This is emphasised by the fact that when using only square basis states the result is considerably higher.\\

\begin{table}[t]
\begin{center}
\begin{tabular}{clll}
\hline
 & Unimproved & Improved & Tadpole Improved   \\
\hline
$\Delta m^S_1$ & $2.05691 \pm 0.00002$ & $ 2.0897\pm 0.0003$ 
& $2.0965\pm 0.0006$ \\
$\Delta m^S_2$ & $3.645 \pm 0.001$ & $ 3.685\pm 0.001$ 
& $3.6953\pm 0.0009$ \\
$\Delta m^S_3$ & $4.5202 \pm 0.0004$ & $4.574\pm 0.004$ 
& $4.583\pm 0.004$ \\
$\Delta m^S_4$ & $5.133\pm 0.003 $ & $5.177\pm 0.004$ 
& $5.189\pm 0.004 $ \\
$\Delta m^S_5$ & $5.867\pm 0.006$ & $5.932\pm 0.008$ 
& $5.943\pm 0.008$ \\
\hline
\end{tabular}
\caption{Upper bounds on the lowest lying SU(2) massgaps (in units of $g^2$) computed with various Hamiltonians in 2+1 dimensions. The unimproved, improved and tadpole results are calculated in the respective scaling regions $13.5 \le \beta \le 30.0$, $9.9 \le \beta \le 30.0$ and $9.25 \le \beta \le 30.0$.} \label{mgg}
\end{center}
\end{table} 

\begin{table}[t]
\begin{center}
\begin{tabular}{clll}
\hline
 & Unimproved & Improved & Tadpole Improved   \\
\hline
$\Delta m^S_1$ &
$3.26506\pm 0.00007 $ 	& $3.3201\pm 0.0003$	& $3.3350\pm 0.0008$ \\
$\Delta m^S_2$ &
$6.194\pm 0.003 $	& $6.273\pm 0.003$	& $6.303\pm 0.002$\\
$\Delta m^S_3$ &
$7.396\pm 0.010 $	& $7.536\pm 0.008$	& $7.575\pm 0.006$\\
$\Delta m^S_4$ &
$8.844\pm 0.007 $	& $8.936\pm 0.008$	& $8.980\pm 0.006$\\
$\Delta m^S_5$ &
$9.87 \pm 0.02  $	& $10.04\pm 0.01 $ 	& $10.10\pm 0.01$\\
\hline
\end{tabular}
\caption{Upper bounds on the lowest lying symmetric SU(3) massgaps (in units of $g^2$) computed with various Hamiltonians in 2+1 dimensions. The results are calculated in the scaling region $18.5 \le \beta \le 50.0$.}\label{mgmg} 
\end{center}
\end{table} 

\begin{table}[t]
\begin{center}
\begin{tabular}{clll}
\hline
 & Unimproved & Improved & Tadpole Improved   \\
\hline
$\Delta m^A_1$ &
$5.281\pm 0.003$ & $5.371\pm 0.002$ & $5.396\pm 0.001$ \\ 
$\Delta m^A_2$ &
$7.831\pm 0.007$ & $7.947\pm 0.006 $& $7.986\pm 0.004 $ \\
$\Delta m^A_3$ &
$7.40\pm 0.01$  & $8.87\pm 0.01 $ & $8.922\pm 0.009 $ \\
$\Delta m^A_4$ &
$10.32\pm 0.01$ & $10.46\pm 0.01$ & $10.514\pm 0.009$ \\
$\Delta m^A_5$ &
$11.06\pm 0.02$ & $11.27\pm 0.02 $ & $11.33\pm 0.01 $ \\
\hline
\end{tabular}
\caption{Upper bounds on the lowest lying antisymmetric SU(3) massgaps (in units of $g^2$) computed with various Hamiltonians in 2+1 dimensions. The results are calculated in the scaling region $18.5 \le \beta \le 50.0$.} \label{mgmgmg}
\end{center}
\end{table}

When compared to equivalent unimproved calculations, the improved and
tadpole improved massgaps approach scaling faster as $\beta$ is
increased. This is evident in \figs{mgconv}{mgcompare} and is expected since,
for an improved calculation one is closer to the continuum limit when
working at a given coupling. However, for each improved calculation
the scaling behaviour is not as precise as the equivalent unimproved
calculation. This is clear from \tabss{mgg}{mgmgmg}. A possible reason for this is that the one plaquette trial state
used here does not allow for direct contributions from the improvement term in
the kinetic Hamiltonian. For this term to contribute directly one would need a
trial state which includes Wilson loops extending at least two links
in at least one direction. \\

The improved SU(2) massgap can be compared to the coupled cluster 
calculation of Li et al~\cite{[22]}. Their result (in units of $g^2$), $\Delta m^S_1 = 1.59$, is again significantly lower than our result $2.0897\pm0.0003$. The difference is attributable to the different choices of ground state used. While our calculation makes use of the simple one plaquette ground state, the coupled cluster 
calculation of Li et al uses a more accurate ground state wave function consisting of an exponential of a sum of extended loops. \\

We can also calculate massgap ratios. 
%in the symmetric sector
%\bea
%	\frac{\Delta m_2^S}{\Delta m_1^S} &=& \Bigg\{
%\begin{array}{lll}
%1.8944 \pm 0.0007 & \quad & \rm{Unimproved} \\
%1.887 \pm 0.001 & \quad & \rm{Improved} \\
%1.887 \pm 0.001 & \quad & \mbox{Tadpole improved} 
%\end{array}
%\eea
% and in the antisymmetric sector
%\bea
%	\frac{\Delta m_2^A}{\Delta m_1^A} &=& \Bigg\{
%\begin{array}{lll}
%1.4815 \pm 0.0001 & \quad & \rm{Unimproved} \\
%1.478 \pm 0.002 & \quad & \rm{Improved} \\
%1.478 \pm 0.002 & \quad & \mbox{Tadpole improved.} 
%\end{array}
%\eea
For example the antisymmetric-symmetric massgap ratio for SU(3) is given by
\bea
	\frac{\Delta m_1^A}{\Delta m_1^S} &=& \Bigg\{
\begin{array}{lll}
1.6173 \pm 0.0008 & \quad & \rm{Unimproved} \\
1.6177 \pm 0.0006 & \quad & \rm{Improved} \\
1.6182 \pm 0.0005 & \quad & \mbox{Tadpole improved.} 
\end{array}
\eea
These results lie between the unimproved results of Chen et al, $1.6989$~\cite{[23]} and Teper, $1.50\pm 0.03$~\cite{[21]}.

\section{Conclusion}

In this paper we have extended the analytic techniques of 2+1
dimensional Hamiltonian LGT, traditionally used for SU(2), to the case
of SU(3).  Impressive scaling is achieved over an extremely wide range
of couplings for the lowest energy eigenstates in the symmetric and
antisymmetric sectors.  Our results use a one plaquette trial state
and a basis of rectangular states over which excited state energies
are minimised.  Such choices allow us to use analytic techniques for
both SU(2) and SU(3).  Our results provide upper bounds on SU(2) and
SU(3) unimproved, improved and tadpole improved glueball masses all of
which are above current estimates.  We expect that these upper bounds
will lower with the use of more complicated states in the simulation
of both the ground state and the excited state.  For such choices of
states the analytic techniques used here are not as readily
applicable.

\section*{Acknowledgements}
The work was supported in part by the Australian Research Council.
\appendix
\section{Commutation Relations}

In improved Hamiltonian LGT calculations, one encounters matrix elements of the form: 
\bea
\langle \phi_0 |\sum_{\boldx,i}\Tr\left[E_i(\boldx)E_i(\boldx)\right]| \phi_0 
\rangle\quad {\rm and} \quad 
\langle \phi_0 |\sum_{\boldx,i}\Tr \left[E_i(\boldx)U_i(\boldx)E_i(\boldx+a\boldsymbol{i})U^\dagger_i(\boldx)
\right]| \phi_0
 \rangle.
\label{me}
\eea
The first of these is easily handled. One simply writes the electric field operators $E_i(\boldx)$ in terms of their components with respect to the Gell-Mann matrices ${\lambda^\alpha}$, $\alpha=1,\ldots,N^2-1$,
\bea
E_i(\boldx) = E^\alpha_i(\boldx)\lambda^\alpha.  
\eea
Making use of the SU$(N)$ relation
\bea
\Tr (\lambda^\alpha \lambda^\beta) = \frac{1}{2} \delta_{\alpha \beta},
\label{trace}
\eea 
we have,
\bea
\langle \phi_0 |\sum_{\boldx,i}\Tr \left[E_i(\boldx)E_i(\boldx)\right]| \phi_0 \rangle 
&=& \frac{1}{2}\sum_{\boldx,i}\langle \phi_0 |E^\alpha_i(\boldx)E^\alpha_i(\boldx)| \phi_0 \rangle.
\eea
Let the trial state $|\phi_0\rangle $ have the form $|\phi_0\rangle = e^S|0\rangle$, where $S$ is a function of link variables and $S^\dagger = S$. Then
\bea
\sum_{\boldx,i}\langle \phi_0| E^\alpha_i(\boldx) E^\alpha_i(\boldx) |\phi_0\rangle = -\sum_{\boldx,i} \langle 0| \left[E^\alpha_i(\boldx), e^S \right]\left[E^\alpha_i(\boldx), e^S\right] |0\rangle .
\label{blerk}
\eea
Making use of the Baker-Hausdorff formula we can derive the following result,
\bea
e^S E^\alpha_i(\boldx) e^{-S} &=& E^\alpha_i(\boldx) - [E^\alpha_i(\boldx),S] + \frac{1}{2}[[E^\alpha_i(\boldx),S],S] 
+\ldots \nn\\
&=& E^\alpha_i(\boldx) - [E^\alpha_i(\boldx),S].
\label{comcom}
\eea
The last line follows from the fact that $S$ is a function of link variables 
and the commutation relation:
\bea
\left[E^\alpha_i(\boldx),U_j(\boldsymbol{y})\right]&=&
\delta_{ij} \delta_{\boldx \boldsymbol{y}}\lambda^\alpha U_{i}(\boldx). 
\label{unmodcom}
\eea
Some rearrangement of \eqn{comcom} leads to the useful result,
\bea
\left[E^\alpha_i(\boldx), e^S\right] = \left[E^\alpha_i(\boldx), S\right] e^S =
e^S \left[E^\alpha_i(\boldx), S\right] ,\label{useful}
\eea 
which we apply in \eqn{blerk} to obtain,
\bea
\sum_{\boldx,i}\langle \phi_0| E^\alpha_i(\boldx) E^\alpha_i(\boldx) |\phi_0\rangle &=& -\sum_{\boldx,i}\langle \phi_0|\left[E^\alpha_i(\boldx), S \right]\left[E^\alpha_i(\boldx), S\right] |\phi_0\rangle \nn\\
&=&  \frac{1}{2}\sum_{\boldx,i} \langle \phi_0| \left[E^\alpha_i(\boldx), 
\left[E^\alpha_i(\boldx),S\right] \right]|\phi_0 \rangle.\label{simplify}
\eea
In the last line we have used integration by parts~\cite{[24]}.\\ 

The second of the matrix elements in \eqn{me} is more difficult to handle.
For this case it is convenient to expand $E_i(\boldx+a\boldsymbol{i})$ in the basis ${\tilde{\lambda}^\alpha_i(\boldx+a\boldsymbol{i})}$:
\bea
E_i(\boldx+a\boldsymbol{i}) = \tilde{\lambda}^\alpha_i(\boldx+a\boldsymbol{i}) \tilde{E}^\alpha_i(\boldx+a\boldsymbol{i}),
\label{crud}
\eea
where,
\bea
\tilde{\lambda}^\alpha_i(\boldx+a\boldsymbol{i}) = U^\dagger_i(\boldx) \lambda^\alpha  U_i(\boldx).  
\label{def}
\eea 
We now derive the commutation relations between $\tilde{E}^\alpha_i(\boldx)$ and $U_{j}(\boldsymbol{y})$. We start with the analogous relation 
to \eqn{trace}, 
\bea
\Tr \left[\tilde{\lambda}^\alpha_i(\boldx) \tilde{\lambda}^\beta_i(\boldx) \right]
= \frac{1}{2} \delta_{\alpha \beta}, \label{trrel}
\eea 
which follows trivially from \eqn{def}.
This relation allows us to invert \eqn{crud} and write $\tilde{E}^\alpha$ in terms of $E^\alpha $
\bea
\tilde{E}^\alpha_i(\boldx) &=& 2 \Tr\left[E_i(\boldx)\tilde {\lambda}^\alpha_i(x)
\right] \nn\\
&=& 2 \Tr\left[\tilde{\lambda}^\alpha_i(\boldx) 
\lambda^\beta \right] E^\beta_i(\boldx).
\label{thatone}
\eea
We can form a similar relation between $\tilde{\lambda}^\alpha$ and $\lambda^\alpha$ as follows. Since $\tilde{\lambda}^\alpha_i(\boldx)\in {\rm SU}(N)$, we can expand in the Gell-Mann basis as follows,
\bea
\tilde{\lambda}^\alpha_i(\boldx) = c^{\alpha \gamma}_i(\boldx) \lambda ^\gamma.
\label{thisone}
\eea
Here $c^{\alpha \beta}_i(x)$ are constants which we now determine.
Multiplying \eqn{thisone} throughout by $\lambda^\beta$, tracing and making use of \eqn{trace} gives 
\bea
c^{\alpha \beta}_i(\boldx) = 
2\Tr\left[\tilde{\lambda}^\alpha_i(\boldx) \lambda ^\beta \right].
\eea
From \eqn{thatone} we see that $\tilde{E}^\alpha_i(\boldx) =  c^{\alpha \beta}_i(\boldx) E^\beta_i(\boldx)$. The commutation relations follow immediately,
\bea
\left[ \tilde{E}^\alpha_i(\boldx),U_{j}(\boldsymbol{y})\right] &=& c^{\alpha \beta}_i(\boldx)
\left[ E^\beta_i(\boldx),U_{j}(\boldsymbol{y})\right] = c^{\alpha \beta}_i(\boldx) \lambda^\beta
\delta_{ij} \delta_{\boldx\boldsymbol{y}}U_{i}(\boldsymbol{x}) \nn\\
&=& \delta_{ij} \delta_{\boldx \boldsymbol{y}}\tilde{\lambda}^\alpha_i(\boldx)U_{i}(\boldx). 
\label{modcom}
\eea
These commutation relations enable the simplification the second of the matrix elements in \eqn{me} using the basis defined in \eqn{crud}. Making use of \eqns{thatone}{modcom}
leads to the following simplification: 
\bea
\!\!\!\!\!\!\!\!\!\!\!
\langle \phi_0 |\sum_{\boldx,i}\Tr \left[E_i(\boldx)U_i(\boldx)E_i(\boldx\!+\!a\boldsymbol{i})U^\dagger_i(\boldx)
\right]| \phi_0 \rangle &\!\!\!=\!\!\!& \frac{1}{4}\sum_{\boldx,i}\langle \phi_0 |
\left[E^\alpha_i(\boldx),[\tilde{E}^\alpha_i(\boldx\!+\!a\boldsymbol{i}),S\right]| \phi_0\rangle .
\label{simplify2}
\eea
In order to calculate such matrix elements the following results are useful
\bea
\left[ E^\alpha_i(\boldx),\left[\tilde{E}^\alpha_i(\boldx+a\boldsymbol{i}),U_i(\boldx)U_i(\boldx+a\boldsymbol{i})\right]\right] 
&\!\!\!=\!\!\!& \frac{N^2-1}{2N}U_i(\boldx)U_i(\boldx+a\boldsymbol{i}),\label{com1}\\
\left[E^\alpha_i(\boldx),\{U_i(\boldx)\}_{AB}\right]\left[\tilde{E}^\alpha_i(\boldx+a\boldsymbol{i}),\{U_i(\boldx+a\boldsymbol{i})\}_{CD}\right]&\!\!\!=\!\!\!& \nn\\
 && \hspace{-7.5cm}\frac{1}{2}\delta_{B C}\{U_i(\boldx)\}_{A B'}
\{U_i(\boldx+a\boldsymbol{i})\}_{B'D} -  \frac{1}{2N}\{U_i(\boldx)\}_{AB}
\{U_i(\boldx+a\boldsymbol{i})\}_{CD}, \\
\left[E^\alpha_i(\boldx),\{U^\dagger_i(\boldx)\}_{AB}\right]\left[\tilde{E}^\alpha_i(\boldx+a\boldsymbol{i}),\{U_i(\boldx+a\boldsymbol{i})\}_{CD}\right]&\!\!\!=\!\!\!& \nn\\
 && \hspace{-7cm}-\frac{1}{2}\{U^\dagger_i(\boldx)\}_{C B}
\{U_i(\boldx+a\boldsymbol{i})\}_{A D} +  \frac{1}{2N}\{U^\dagger_i(\boldx)\}_{AB}
\{U_i(\boldx+a\boldsymbol{i})\}_{CD},\\
\left[E^\alpha_i(\boldx),\{U_i(\boldx)\}_{AB}\right]\left[\tilde{E}^\alpha_i(\boldx+a\boldsymbol{i}),\{U^\dagger_i(\boldx+a\boldsymbol{i})\}_{CD}\right]&\!\!\!=\!\!\!& \nn\\
 && \hspace{-7cm}-\frac{1}{2}\{U_i(\boldx)\}_{C B}
\{U^\dagger_i(\boldx+a\boldsymbol{i})\}_{AD} +  
\frac{1}{2N}\{U_i(\boldx)\}_{AB}
\{U^\dagger_i(\boldx+a\boldsymbol{i})\}_{CD}.\label{com4}
\eea
Here $\{X\}_{AB}$ denote the colour indices of the matrix $X$ and an implicit sum over repeated colour indices is understood. 
These results follow simply from the commutation relations of 
\eqns{unmodcom}{modcom}, and the following SU($N$) formula:
\bea
\lambda_{AB}^\alpha \lambda_{CD}^\alpha = 
\frac{1}{2}\delta_{AD}\delta_{BC} - \frac{1}{2N} \delta_{AB}\delta_{CD}.
\eea

\end{document}